%% file: main.tex
\documentclass[conference]{IEEEtran}
\usepackage{etoolbox}
\IEEEoverridecommandlockouts
\usepackage{listings}
\usepackage{cite}
\usepackage{dsfont}
\usepackage{amsmath, amssymb}
\usepackage{tikz}
\usepackage{pgfplots}
\pgfplotsset{compat=1.16}
\usepackage{adjustbox}
\usepackage{xcolor, soul}
\usetikzlibrary{backgrounds,calc,shadings,shapes,arrows,arrows.meta,shapes.arrows,shapes.symbols,patterns, datavisualization, datavisualization.formats.functions, matrix}
\usepackage[vlined,ruled]{algorithm2e}

\AfterEndPreamble{
  \mathchardef\mathcomma\mathcode`\,
  \mathcode`\,="8000 
}
{\catcode`,=\active
  \gdef,{\mathcomma\discretionary{}{}{}}
}

\allowdisplaybreaks %in order to split big formulas across different breaks, e.g., pages, or columns

\begin{document}

\title{On using SMT-solvers for Modeling and Verifying Dynamic Network Emulators\\\large{(Work in Progress)}
}

\author{\IEEEauthorblockN{Erick Petersen$^{1,2}$, Jorge L\'opez$^1$, Natalia Kushik$^2$, Claude Poletti$^1$, and Djamal Zeghlache$^2$
\IEEEauthorblockA{%\textit{dept. name of organization (of Aff.)} \\
$^1$ \textit{Airbus Defence and Space}, \'Elancourt, France \\
%\{jorge.lopez-c, claude.poletti\}@airbus.com\\
$^2$ \textit{T\'{e}l\'{e}com SudParis, Institut Polytechnique de Paris}, Palaiseau, France\\ %\{erick\_petersen, natalia.kushik, djamal.zeghlache\}@telecom-sudparis.eu
}}}
\maketitle

\begin{abstract}
\input{abstract}
\end{abstract}

\section{Introduction}\label{sec:intro}
\input{introduction}

\section{Modeling networks using MSFOL}\label{sec:verif}
\input{modeling}

\section{Model checking \& run-time verification for dynamic network emulators}\label{sec:run-time_verification}
\input{verification}
\input{experiments}

\section{Conclusion}\label{sec:conc}
\input{conclusion}

\bibliographystyle{IEEETran}
\bibliography{references}

\end{document}

%% file: abstract.tex
A novel model-based approach to verify dynamic networks is proposed; the approach consists in formally describing the network topology and dynamic link parameters. A many sorted first order logic formula is constructed to check the model with respect to a set of properties. The network consistency is verified using an SMT-solver, and the formula is used for the run-time network verification when a given static network instance is implemented. The z3 solver is used for this purpose and corresponding preliminary experiments showcase the expressiveness and current limitations of the proposed approach.

%% file: introduction.tex
As dynamic networks progress rapidly, novel solutions emerge. Testing and evaluating network components in a well emulated environment \cite{lai2020network} is necessary to reduce risks and save time at the final deployment. We focus on \emph{dynamic} networks whose parameters change, e.g., the bandwidth and delay of wireless links may change due to external interference. Depending on the desired network to emulate, the link parameters vary according to properties that are to be respected by the emulators (e.g., large delays on distant objects).

This paper is devoted to the problem of adequate emulation of such networks. We propose to rely on model checking and run-time verification techniques that under certain assumptions allow checking that the emulator is adequate for a given context and is designed correctly. The novelty of the proposed approach for the emulator design and verification relies on the use of Many Sorted First Order Logic~(MSFOL) for describing the dynamic network characteristics. The latter allows to automatically validate if the network description is consistent, and after the network is implemented, to verify at run-time that it behaves as requested. Both tasks can be effectively implemented through the use of a Satisfiability Modulo Theories~(SMT) solver. Note that existing network simulators and emulators (e.g., \cite{priyadarshi2020deployment}) provide a solid background, however the authors are not aware of any works related to model-based dynamic network verification. We also note that the network description for a network emulator is another challenging issue. Existing solutions can provide convenient interfaces \cite{varga2001discrete} for network description or allow specific scenario files (e.g., \cite{riley2010ns}). However, such descriptions remain rather informal and do not facilitate further verification. 

We propose to utilize a formal network description that contains both, the properties of the network topology as well as various network parameters. As mentioned before, such description is provided in terms of a MSFOL formula to be verified by an SMT-solver; in this paper, the z3 solver \cite{z3} is used. Preliminary experiments showcase the flexibility provided by this approach. Likewise, we explore the performance for large network instances and discuss the current limitations of the formal verification for dynamic networks.

%% file: modeling.tex
\begin{figure}
    \centering
    \input{figures/example_net_topo}
    \caption{Example dynamic network}
    \label{fig:ex_dyn_net}
\end{figure}
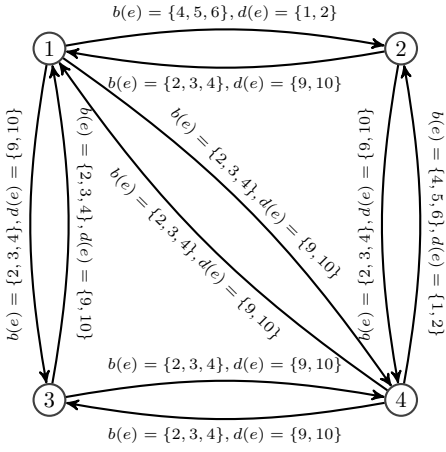

\textbf{Dynamic networks.} A static network is a computer network where each link has a set of parameters that do not change, for example bandwidth (capacity) or delay. Differently from static networks, the parameters of the links may change in dynamic networks (in the scope of our current work, we assume the network topology does not change); such change can be the consequence of the physical medium (e.g., in wireless / radio frequency networks) or due to logical changes (e.g., rate limiting the capacity of a given link). Static networks can be modeled as (directed) weighted graphs $(V, E, p_1,\ldots,p_k)$, where $V$ is a set of nodes, $E\subseteq V\times V$ is a set of directed edges, and $p_i$ is a link parameter function $p_i: E \to \mathds{N}$, for $i\in\{1,\ldots,k\}$; without loss of generality, we assume that the parameter functions map to non-negative integers (denoted by $\mathds{N}$) or related values can be encoded with them. Similarly, dynamic networks can be modeled as such graphs, however, $p_i$ maps an edge to a non-empty set of integer values, i.e., $p_i: E \to 2^{\mathds{N}}\setminus\emptyset$, where $2^{\mathds{N}}$ denotes the power-set of $\mathds{N}$. As an example, consider the dynamic network depicted in Fig.~\ref{fig:ex_dyn_net}, and its model $\mathcal{N}=(V,E,p_1(e),p_2(e))$, where:
\[
\begin{footnotesize}
\begin{aligned}
V&=\{1,2,3,4\}\\
E&=\{(1,2),(2,1),(1,3),(3,1),(1,4),(4,1),(2,4),(4,2),(3,4)(4,3)\}\\
p_1(e)&=b((s,d))=\begin{cases}
        \{4,5,6\}, & \text{if } d = 2\\
        \{2,3,4\}, & \text{otherwise}
        \end{cases}\\ 
p_2(e)&=d((s,d))=\begin{cases}
        \{1,2\}, & \text{if } d = 2\\
        \{9,10\}, & \text{otherwise}
        \end{cases}.
\end{aligned}    
\end{footnotesize}
\]

Semantically, this model represents a dynamic network in which the link's available bandwidth can vary according to the function $b$ (for \emph{bandwidth}), and the link's delay can vary according to the function $d$ (for \emph{delay}). Note that a dynamic network snapshot, at a given time instance, is a static network, and thus, we use both terms interchangeably.

\textbf{Satisfiability Modulo Theories~(SMT).} SMT is concerned with the satisfiability of formulas with respect to some background theory \cite{smt1}. SMT usually works with a typed version of first order logic, particularly MSFOL. The syntax for MSFOL formulas is standard (see \cite{smt1}), however, in our work, we use $x:\sigma$ to indicate that $x$ is of sort (type) $\sigma$ (e.g., in quantified formulas $\forall x:\sigma\; \phi$). We denote as $f: \sigma_1\times\sigma_2\times\ldots\times\sigma_n\to\sigma$ that function $f$ is declared of sort $\sigma$.

\textbf{Modeling dynamic networks using MSFOL formulas.} It is desirable that our model uses sorts of theories which are available in SMT solvers and ideally those which are decidable. For that reason, (finite) sets are encoded as objects of array sort (denoted $\mathcal{A}$), e.g., the set $V=\{1,2,3,4\}$ is encoded as $\phi_V=(V:\mathcal{A_{\mathds{Z},\mathds{Z}}})\wedge(V[1]=1)\wedge(V[2]=2)\wedge(V[3]=3)\wedge(V[4]=4)\wedge(|V|=4)$. For convenience, we specify that the sort of $V$ is an array whose indices and values are integers; likewise, we also include the cardinality of $V$. Directed edges are nothing more than records (tuples with sorts), particularly, pairs of integers. Thus, following our encoding, $\phi_E=(E:\mathcal{A_{\mathds{Z},\mathds{Z}\times\mathds{Z}}})\wedge(E[1]=(1,2))\wedge\ldots\wedge(E[10]=(4,3))\wedge(|E|=10)$. In order to model the bandwidth of an edge ($b(e)$) and the delay of an edge ($d(e)$) according to the given functions, we use the formula $\phi_{p}=(b:\mathds{Z}\times \mathds{Z}\to\mathds{Z})\wedge(d:\mathds{Z}\times \mathds{Z}\to\mathds{Z})\wedge\forall x:\mathds{Z}\;\;( ((x\geq 1)\wedge(x\leq |E|)) \implies 
(   
    (
        (dst(E[x])=2)
        \implies 
        ( 
            (b(E[x])\geq 4) \wedge 
            (b(E[x])\leq 6) \wedge 
            (d(E[x])\geq 1) \wedge 
            (d(E[x]\leq 2))
        )
    )
    \wedge
    (
        (dst(E[x])\not=2)
        \implies 
        ( 
            (b(E[x])\geq 2) \wedge 
            (b(E[x])\leq 4) \wedge 
            (d(E[x])\geq 9) \wedge 
            (d(E[x]\leq 10))
        )
    )
) )$. The complete model is the conjunction of the three previous formulas, i.e., $\phi_\mathcal{N}=\phi_V\wedge\phi_E\wedge\phi_{p}$. This simple example showcases very little of the flexibility provided by describing dynamic networks as MSFOL formulas, however very complex models can be described with such formalism.

%% file: figures/example_net_topo.tex
\begin{tikzpicture}[node distance=5.5cm,>=stealth',bend angle=45,auto,scale=.9, every node/.style={scale=.85}]
    \tikzstyle{switch}=[circle,thick,draw=black!75, inner sep=0.075cm]
    \tikzstyle{host}=[circle,thick,draw=black!75,fill=black!80,text=white, inner sep=0.075cm]
    \tikzstyle{undirected}=[thick]
    \tikzstyle{directed}=[thick,->]

    \node[switch] (s1) {$1$};
    \node[switch, right of=s1] (s2) {$2$};
    \node[switch, below of=s1] (s3) {$3$};
    \node[switch, below of=s2] (s4) {$4$};
   
    \path   
            (s1)    edge[directed, bend left=10]    node[above] {\scriptsize $b(e)=\{4,5,6\}, d(e)=\{1,2\}$}   (s2)
                    edge[directed, bend right=10]    node[above, rotate=90] {\scriptsize $b(e)=\{2,3,4\}, d(e)=\{9,10\}$}   (s3)
                    edge[directed, bend left=10]    node[above, rotate=-45] {\scriptsize $b(e)=\{2,3,4\}, d(e)=\{9,10\}$}   (s4)
            (s2)    edge[directed, bend left=10]    node[below] {\scriptsize $b(e)=\{2,3,4\}, d(e)=\{9,10\}$}  (s1)
                    %edge[directed, bend right=10]    node[above, rotate=45] {$b(e)=\{2,3,4\}, d(e)=\{9,10\}$}   (s3)
                    edge[directed, bend right=10]    node[above, rotate=90] {\scriptsize $b(e)=\{2,3,4\}, d(e)=\{9,10\}$}   (s4)
            (s3)    edge[directed, bend right=10]    node[above, rotate=-90] {\scriptsize $b(e)=\{2,3,4\}, d(e)=\{9,10\}$}   (s1)
                    %edge[directed, bend right=10]    node[below, rotate=45] {$b(e)=\{4,5,6\}, d(e)=\{1,2\}$}   (s2)
                    edge[directed, bend left=10]    node[above] {\scriptsize $b(e)=\{2,3,4\}, d(e)=\{9,10\}$}   (s4)
            (s4)    edge[directed, bend left=10]    node[below, rotate=-45] {\scriptsize $b(e)=\{2,3,4\}, d(e)=\{9,10\}$}   (s1)
                    edge[directed, bend right=10]    node[above, rotate=-90] {\scriptsize $b(e)=\{4,5,6\}, d(e)=\{1,2\}$}   (s2)
                    edge[directed, bend left=10]    node[below] {\scriptsize $b(e)=\{2,3,4\}, d(e)=\{9,10\}$}   (s3);
\end{tikzpicture}

%% file: verification.tex
\textbf{Model checking.} Once the model of a dynamic network is built, a set of properties of interest can be verified. For example, checking that the values of the parameters are not negative can be easily expressed with the MSFOL formula: $\pi_> = \forall x:\mathds{Z}\;\; ( ((x\geq 1)\wedge(x\leq |E|)) \implies ((b(e) \geq 0) \wedge (d(e) \geq 0) ) )$. Likewise, checking that all nodes have at least one incoming and one outgoing edge can be easily expressed as: $\pi_{\downarrow\uparrow} = \forall x:\mathds{Z}\;\; ( ((x\geq 1)\wedge(x\leq |V|)) \implies \exists y,z:\mathds{Z}\;\; ((y \not= z) \wedge (y \geq 1) \wedge (y \leq |E|) \wedge (z \geq 1) \wedge (z \leq |E|) \wedge (src(E[y])=V[x]) \wedge (dst(E[z])=V[x]) )  )$.  

Having the properties to be checked, $\pi\mathcal{C}$ as a conjunction of them (in our previous example $\pi\mathcal{C}=\pi_>\wedge\pi_{\downarrow\uparrow}$), and the model representing the dynamic network $\phi_\mathcal{N}$, the model checking process is quite straightforward. First, we check that both $\phi_\mathcal{N}$ and $\pi_\mathcal{C}$ are satisfiable; otherwise, either the model (whose verification can be performed beforehand as well) or the properties have inconsistencies. Further, if the formula $\phi_\mathcal{N}\wedge\pi_\mathcal{C}$ is satisfiable, then we conclude that \emph{the properties $\pi_\mathcal{C}$ are held for the model $\phi_\mathcal{N}$}. If the formula is not satisfiable it implies that there does not exist a satisfiable interpretation for both formulas at the same time, i.e., that there is a conflict between $\phi_\mathcal{N}$ and $\pi_\mathcal{C}$. 

As an example, consider the formula $\phi_\mathcal{N}$ associated with the dynamic network shown in Figure~\ref{fig:ex_dyn_net}, additionally, consider the property $\pi_{d>2} = \forall x:\mathds{Z}\;\; ( ((x\geq 1)\wedge(x\leq |E|)) \implies ((d(e) > 2) ) )$ stating that the delay must be at least equal to three. $\phi_\mathcal{N}\wedge\pi_{d<2}$ is not satisfiable, as the model states that the delay of certain edges is either one or two, and thus, there is no satisfiable interpretation for the conjunction of formulas (even if there are satisfiable interpretations for each of them). On the contrary, $\phi_\mathcal{N}\wedge\pi_{>}\wedge\pi_{\downarrow\uparrow}$ is satisfiable.

\textbf{Run-time verification.} Once a dynamic network emulator is implemented, it can be continuously verified that the produced static instances do not violate the description of $\mathcal{N}$. In this case, we consider a conformance relation $\preceq$ which is similar to a \emph{reduction}, i.e., a static network $\mathcal{N_S} \preceq \mathcal{N}$ if the topology $(V, E)$ of $\mathcal{N_S}$ is exactly the same of that one of $\mathcal{N}$, and for each $i \in \{1, \dots, k\}$ $p_{i_{\mathcal{N_S}}}(e) \in p_{i_\mathcal{N}}(e)$. 

In order to verify this relation $\preceq$ at run-time, the behavior of the emulator can be monitored for checking that each link $(v_a, v_b) \in V \times V$ is implemented correctly, and that each value of the $i$-th parameter belongs to the set $p_{i_\mathcal{N}}((v_a, v_b))$. We assume that the points of observation in this case can be placed at each node $v \in V$ and each link $(v_a, v_b)$, correspondingly. We propose to iteratively check the implementation of each link verifying that the value of $p_{i_{\mathcal{N_S}}}((v_a,v_b))$ does not violate the description of the dynamic network $\mathcal{N}$. The latter can be performed through a call to an SMT-solver and whenever the corresponding formula is not satisfiable, an alert is produced. When alerting, relevant information for debugging can be shown, i.e., the link $(v_1, v_2)$ itself as well as the $i$-th parameter which was wrongly assigned when implementing the static instance $\mathcal{N_S}$. The corresponding procedure is shown in Algorithm~\ref{algo:monitoring}.

\begin{algorithm}[!ht]
    \footnotesize
    \SetKwInOut{Input}{input}\SetKwInOut{Output}{output}\SetKw{KwBy}{by}
    \Input{A formula $\phi_\mathcal{N}$ specifying the dynamic network $\mathcal{N}$}
    \Output{Alert for a static network violating $\preceq$; a link and a parameter `responsible' for the violation}
    
    \While{$true$} 
    {%\tcp{$working$ is a Boolean flag to control the execution of the monitoring process}
        Get the dynamic network instance (static network) $\mathcal{N_S}$\;
        \ForEach{link $(v_a, v_b)$}
        {
            create a formula $\phi=\exists x:\mathds{Z}\;\;((x\geq1)\wedge(x\leq|E|)\wedge(E[x]=(v_a, v_b)))$\;
            \If{$\phi_\mathcal{N} \wedge \phi$ is $UNSAT$} 
            {
                $alert((v_a, v_b))$\;%\tcp{Alert a wrong implementation of link $(v_a, v_b)$}
            }%if
            $\phi_\mathcal{N} = \phi_\mathcal{N} \wedge \phi$\;
            \ForEach{$i\in\{1, \dots, k\}$}
            {create a formula $\phi_i=(p_{i_\mathcal{N}}((v_a,v_b))=p_{i_{\mathcal{N_S}}}((v_a,v_b)))$\;
                \If{$\phi_\mathcal{N} \wedge \phi_i$ is $UNSAT$} 
                {
                    $alert(i, (v_a,v_b))$\;%\tcp{Alert a wrong assignment of the $i$-th parameter on link $(v_a,v_b)$}
                }%if
        
            }
        }%for each

    }%while
    \caption{Run-time verification of a dynamic network emulator}\label{algo:monitoring}
\end{algorithm}

%% file: experiments.tex
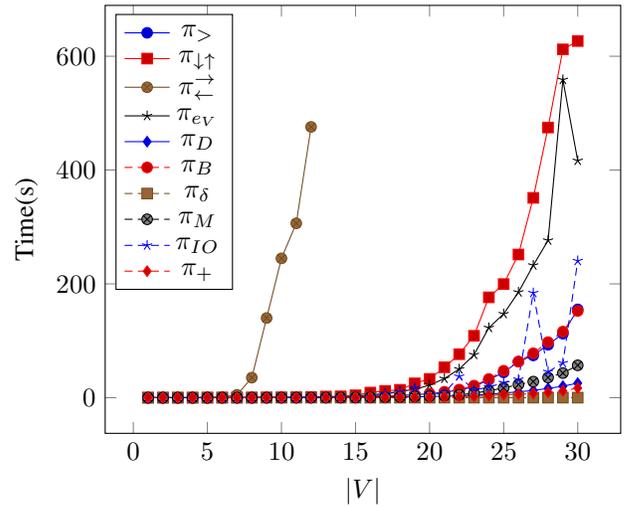
\begin{figure}
    \centering
    \input{figures/time_eval}
    \caption{Time evaluation of emulator verification}
    \label{fig:time}
\end{figure}

\textbf{Preliminary experiments.} In order to showcase the expressiveness of the proposed method, and to assess its limitations, a preliminary experimental evaluation has been conducted. From the model checking point of view, the properties $\pi_>$ and $\pi_{\downarrow\uparrow}$, as previously described, have been tested alongside the properties listed in Table~\ref{tab:props}; specific functions to compute the bandwidth ($in, out$, and $bw$) are not described in order to avoid overloading the formulas. The properties have been coded in SMT-LIB and the z3 solver has been used to check their satisfiability in randomly generated graphs. The running time for networks ranging from one to 30 nodes is shown in Fig.~\ref{fig:time}; note that large instances can require long verification time. We then conclude that such method can be rather inefficient for run-time verification when the whole network is verified at once. For a large network of 50 nodes, a somewhat complex property as $\pi_{IO}$ can be verified in approximately 12 minutes. If verifying a simple edge conformance as proposed in Algorithm~\ref{algo:monitoring}, the time decreases to five minutes; this is one of the reasons why we propose such an incremental approach. However, the best results have been obtained when a \emph{backtracking} of the variable of interest is done, and only the relevant variables that are involved in the property are kept. Even for a large network of 50 nodes, the verification time per edge is around 0.069s. For this work in progress, this process has been manually performed; an automated process is envisioned for future work. Note that all z3 code, graphs and properties can be found in our repository \cite{repo}.

\begin{table}[!htb]
    \centering
    \begin{tabular}{|p{4cm}|p{4.5cm}|}
    \hline
         \textbf{Description} & \textbf{Formula} \\ \hline
         The links are symmetric (for any link a return link exists) & $\pi_{\leftarrow}^{\rightarrow} = \forall x:\mathds{Z}\;\; ( ((x\geq 1)\wedge(x\leq |E|)) \implies \exists y:\mathds{Z}\;\;  ( (y \geq 1) \wedge (y \leq |E|) \wedge (src(E[x]) = dst(E[y])) \wedge (dst(E[x]) = src(E[y])) )$ \\ \hline
         
         The edges in the edge array are composed of nodes in the node array &  $\pi_{e_V} = \forall i:\mathds{Z}\;\; ( ( (i \geq 1) \wedge  (i \leq |E|) ) \implies (\exists j,k:\mathds{Z}\;\; ( (src(E[i]) = V[j]) \wedge (dst (E[i]) = V[k])  ) ) )$\\ \hline
         
         The delay of all links is always less or equal to a given constant $D$ & $\pi_{D} = \forall i:\mathds{Z}\;\; ( ( (i \geq 1) \wedge  (i \leq |E|) ) \implies (d(E[i]) \leq D) )$ \\ \hline
         
         The bandwidth of all links is greater or equal to the threshold $B$ & $\pi_{B} = \forall i:\mathds{Z}\;\; ( ( (i \geq 1) \wedge  (i \leq |E|) ) \implies (b(E[i]) \geq B) )$ \\ \hline
         
         The network topology density is at least $\delta$ & $\pi_{\delta} = (|E| /  (|V| * (|V| - 1))) \geq \delta$ \\ \hline
         
         The network topology cannot be full mesh& $\pi_{M} = (|E| /  (|V| * (|V| - 1))) \not= 1$ \\ \hline
         
         The incoming bandwidth of all nodes is strictly less than $C$ times the outgoing bandwidth capacity & $\pi_{IO} = (in:\mathds{Z}\to\mathds{Z}) \wedge (out:\mathds{Z}\to\mathds{Z}) \wedge (\forall i:\mathds{Z}\;\; ( ( (i \geq 1) \wedge  (i \leq |V|) ) \implies (in (V[i]) < C* out(V[i])))$ \\ \hline
         
         The sum of the bandwidth of all links cannot exceed the threshold $\mathcal{B}$ & $\pi_{+} = (bw:\mathcal{A}_{\mathds{Z},\mathds{Z}}\to\mathds{Z}) \wedge (bw (E) \leq \mathcal{B})$ \\ \hline
    \end{tabular}
    \vspace{0cm}
    \caption{Network properties of interest}
    \label{tab:props}
\end{table}

%\vspace{-0.8em}
%$\pi_{v}= \nexists i,j:\mathds{Z}\;\; ((i \not= j) \wedge (i \geq 1) \wedge  (i \leq |V|) \wedge (j \geq 1) \wedge  (j \leq |V|) \wedge (V[i]) = V[j]))$

%$\pi_{e}= \nexists i,j:\mathds{Z}\;\; ((i \not= j) \wedge (i \geq 1) \wedge  (i \leq |E|) \wedge (j \geq 1) \wedge  (j \leq |E|) \wedge (E[i]) = E[j]))$

%% file: figures/time_eval.tex
\begin{tikzpicture}
\begin{axis}[
	xlabel={$|V|$},
	ylabel={Time(s)},
	legend style={at={(0.02,0.98)}, anchor=north west},
]
\addplot coordinates 
{
( 1 , 0.057 ) %- unsat
( 2 , 0.041 ) %- sat
( 3 , 0.042 ) %- sat
( 4 , 0.048 ) %- sat
( 5 , 0.044 ) %- sat
( 6 , 0.056 ) %- sat
( 7 , 0.080 ) %- sat
( 8 , 0.116 ) %- sat
( 9 , 0.137 ) %- sat
( 10 , 0.208 ) %- sat
( 11 , 0.252 ) %- sat
( 12 , 0.316 ) %- sat
( 13 , 0.434 ) %- sat
( 14 , 0.537 ) %- sat
( 15 , 0.962 ) %- sat
( 16 , 1.869 ) %- sat
( 17 , 2.471 ) %- sat
( 18 , 3.627 ) %- sat
( 19 , 5.181 ) %- sat
( 20 , 6.888 ) %- sat
( 21 , 9.708 ) %- sat
( 22 , 13.174 ) %- sat
( 23 , 19.522 ) %- sat
( 24 , 31.511 ) %- sat
( 25 , 44.510 ) %- sat
( 26 , 63.286 ) %- sat
( 27 , 74.270 ) %- sat
( 28 , 93.038 ) %- sat
( 29 , 112.446 ) %- sat
( 30 , 155.141 ) %- sat
};\addlegendentry{$\pi_> $}

\addplot coordinates
{

( 1 , 0.041 ) %- unsat
( 2 , 0.048 ) %- sat
( 3 , 0.055 ) %- sat
( 4 , 0.056 ) %- sat
( 5 , 0.063 ) %- sat
( 6 , 0.099 ) %- sat
( 7 , 0.153 ) %- sat
( 8 , 0.221 ) %- sat
( 9 , 0.406 ) %- sat
( 10 , 0.530 ) %- sat
( 11 , 0.796 ) %- sat
( 12 , 1.083 ) %- sat
( 13 , 2.011 ) %- sat
( 14 , 2.378 ) %- sat
( 15 , 4.229 ) %- sat
( 16 , 8.901 ) %- sat
( 17 , 11.889 ) %- sat
( 18 , 13.835 ) %- sat
( 19 , 25.193 ) %- sat
( 20 , 32.939 ) %- sat
( 21 , 53.392 ) %- sat
( 22 , 76.392 ) %- sat
( 23 , 108.795 ) %- sat
( 24 , 176.329 ) %- sat
( 25 , 199.667 ) %- sat
( 26 , 251.533 ) %- sat
( 27 , 351.143 ) %- sat
( 28 , 474.736 ) %- sat
( 29 , 612.073 ) %- sat
( 30 , 626.688 ) %- sat

};\addlegendentry{$\pi_{\downarrow\uparrow} $}

\addplot coordinates
{
( 1 , .042 ) %- unsat
( 2 , .049 ) %- sat
( 3 , .065 ) %- sat
( 4 , .104 ) %- sat
( 5 , .112 ) %- sat
( 6 , 1.196 ) %- sat
( 7 , 4.895 ) %- sat
( 8 , 35.200 ) %- sat
( 9 , 140.011 ) %- sat
( 10 , 244.795 ) %- sat
( 11 , 306.384 ) %- sat
( 12 , 475.828 ) %- sat
%( 13 , 747.047 ) %- sat
%( 14 , 1118.798 ) %- sat
%( 15 , 2637.208 ) %- sat
%( 16 , 11041.454 ) %- sat
%( 17 , 24782.771 ) %- sat
};\addlegendentry{$\pi_{\leftarrow}^{\rightarrow}$}

\addplot coordinates
{
( 1 , .041 ) %- unsat
( 2 , .038 ) %- sat
( 3 , .040 ) %- sat
( 4 , .028 ) %- sat
( 5 , .029 ) %- sat
( 6 , .081 ) %- sat
( 7 , .111 ) %- sat
( 8 , .184 ) %- sat
( 9 , .339 ) %- sat
( 10 , .404 ) %- sat
( 11 , .526 ) %- sat
( 12 , .642 ) %- sat
( 13 , 1.121 ) %- sat
( 14 , 1.448 ) %- sat
( 15 , 2.373 ) %- sat
( 16 , 5.169 ) %- sat
( 17 , 7.182 ) %- sat
( 18 , 11.112 ) %- sat
( 19 , 14.618 ) %- sat
( 20 , 22.104 ) %- sat
( 21 , 33.547 ) %- sat
( 22 , 50.200 ) %- sat
( 23 , 75.420 ) %- sat
( 24 , 122.899 ) %- sat
( 25 , 147.148 ) %- sat
( 26 , 185.342 ) %- sat
( 27 , 232.587 ) %- sat
( 28 , 276.383 ) %- sat
( 29 , 558.707 ) %- sat
( 30 , 416.535 ) %- sat

};\addlegendentry{$\pi_{e_V} $}

\addplot coordinates
{
( 1 , .056 ) %- unsat
( 2 , .038 ) %- unsat
( 3 , .041 ) %- unsat
( 4 , .035 ) %- unsat
( 5 , .061 ) %- unsat
( 6 , .056 ) %- unsat
( 7 , .067 ) %- unsat
( 8 , .064 ) %- unsat
( 9 , .108 ) %- unsat
( 10 , .146 ) %- unsat
( 11 , .112 ) %- unsat
( 12 , .169 ) %- unsat
( 13 , .203 ) %- unsat
( 14 , .240 ) %- unsat
( 15 , .369 ) %- unsat
( 16 , .667 ) %- unsat
( 17 , .869 ) %- unsat
( 18 , 1.199 ) %- unsat
( 19 , 1.600 ) %- unsat
( 20 , 1.978 ) %- unsat
( 21 , 2.543 ) %- unsat
( 22 , 3.278 ) %- unsat
( 23 , 4.457 ) %- unsat
( 24 , 5.915 ) %- unsat
( 25 , 8.129 ) %- unsat
( 26 , 10.303 ) %- unsat
( 27 , 12.738 ) %- unsat
( 28 , 16.261 ) %- unsat
( 29 , 20.167 ) %- unsat
( 30 , 25.506 ) %- unsat
};\addlegendentry{$\pi_{D} $}

\addplot coordinates
{
( 1 , .046 ) %- unsat
( 2 , .036 ) %- sat
( 3 , .046 ) %- sat
( 4 , .039 ) %- sat
( 5 , .057 ) %- sat
( 6 , .073 ) %- sat
( 7 , .086 ) %- sat
( 8 , .115 ) %- sat
( 9 , .176 ) %- sat
( 10 , .175 ) %- sat
( 11 , .273 ) %- sat
( 12 , .344 ) %- sat
( 13 , .503 ) %- sat
( 14 , .617 ) %- sat
( 15 , .977 ) %- sat
( 16 , 1.987 ) %- sat
( 17 , 2.592 ) %- sat
( 18 , 3.857 ) %- sat
( 19 , 5.335 ) %- sat
( 20 , 7.149 ) %- sat
( 21 , 10.331 ) %- sat
( 22 , 14.245 ) %- sat
( 23 , 21.191 ) %- sat
( 24 , 32.845 ) %- sat
( 25 , 47.371 ) %- sat
( 26 , 63.423 ) %- sat
( 27 , 78.060 ) %- sat
( 28 , 97.714 ) %- sat
( 29 , 116.166 ) %- sat
( 30 , 152.350 ) %- sat

};\addlegendentry{$\pi_{B}$}

\addplot coordinates
{
( 1 , .042 ) %- unsat
( 2 , .045 ) %- sat
( 3 , .037 ) %- unsat
( 4 , .036 ) %- unsat
( 5 , .039 ) %- unsat
( 6 , .041 ) %- unsat
( 7 , .036 ) %- unsat
( 8 , .038 ) %- unsat
( 9 , .031 ) %- unsat
( 10 , .037 ) %- unsat
( 11 , .037 ) %- unsat
( 12 , .042 ) %- unsat
( 13 , .039 ) %- unsat
( 14 , .035 ) %- unsat
( 15 , .032 ) %- unsat
( 16 , .028 ) %- unsat
( 17 , .042 ) %- unsat
( 18 , .044 ) %- unsat
( 19 , .051 ) %- unsat
( 20 , .047 ) %- unsat
( 21 , .054 ) %- unsat
( 22 , .056 ) %- unsat
( 23 , .050 ) %- unsat
( 24 , .048 ) %- unsat
( 25 , .030 ) %- unsat
( 26 , .028 ) %- unsat
( 27 , .054 ) %- unsat
( 28 , .053 ) %- unsat
( 29 , .057 ) %- unsat
( 30 , .067 ) %- unsat
};\addlegendentry{$\pi_{\delta}$}

\addplot coordinates
{
( 1 , .044 ) %- unsat
( 2 , .035 ) %- unsat
( 3 , .036 ) %- sat
( 4 , .042 ) %- sat
( 5 , .045 ) %- sat
( 6 , .040 ) %- sat
( 7 , .055 ) %- sat
( 8 , .068 ) %- sat
( 9 , .070 ) %- sat
( 10 , .128 ) %- sat
( 11 , .144 ) %- sat
( 12 , .151 ) %- sat
( 13 , .189 ) %- sat
( 14 , .243 ) %- sat
( 15 , .349 ) %- sat
( 16 , .747 ) %- sat
( 17 , 1.009 ) %- sat
( 18 , 1.432 ) %- sat
( 19 , 2.064 ) %- sat
( 20 , 2.611 ) %- sat
( 21 , 3.615 ) %- sat
( 22 , 4.947 ) %- sat
( 23 , 7.849 ) %- sat
( 24 , 12.079 ) %- sat
( 25 , 17.057 ) %- sat
( 26 , 22.989 ) %- sat
( 27 , 28.195 ) %- sat
( 28 , 35.428 ) %- sat
( 29 , 43.208 ) %- sat
( 30 , 56.855 ) %- sat
};\addlegendentry{$\pi_{M}$}

\addplot coordinates
{
( 1 , .051 ) %- unsat
( 2 , .046 ) %- unsat
( 3 , .050 ) %- unsat
( 4 , .058 ) %- unsat
( 5 , .070 ) %- unsat
( 6 , .121 ) %- unsat
( 7 , .135 ) %- unsat
( 8 , .188 ) %- unsat
( 9 , .291 ) %- unsat
( 10 , .271 ) %- unsat
( 11 , .346 ) %- unsat
( 12 , .943 ) %- unsat
( 13 , 1.764 ) %- unsat
( 14 , .511 ) %- unsat
( 15 , .947 ) %- unsat
( 16 , 1.484 ) %- unsat
( 17 , 6.683 ) %- unsat
( 18 , 10.270 ) %- unsat
( 19 , 17.411 ) %- unsat
( 20 , 4.831 ) %- unsat
( 21 , 8.353 ) %- unsat
( 22 , 37.377 ) %- unsat
( 23 , 13.329 ) %- unsat
( 24 , 18.770 ) %- unsat
( 25 , 26.234 ) %- unsat
( 26 , 32.098 ) %- unsat
( 27 , 184.131 ) %- unsat
( 28 , 45.496 ) %- unsat
( 29 , 60.972 ) %- unsat
( 30 , 240.487 ) %- unsat
};\addlegendentry{$\pi_{IO}$}

\addplot coordinates
{
( 1 , .041 ) %- unsat
( 2 , .040 ) %- sat
( 3 , .039 ) %- sat
( 4 , .043 ) %- unsat
( 5 , .041 ) %- unsat
( 6 , .053 ) %- unsat
( 7 , .053 ) %- unsat
( 8 , .052 ) %- unsat
( 9 , .081 ) %- unsat
( 10 , .088 ) %- unsat
( 11 , .065 ) %- unsat
( 12 , .114 ) %- unsat
( 13 , .094 ) %- unsat
( 14 , .185 ) %- unsat
( 15 , .207 ) %- unsat
( 16 , .280 ) %- unsat
( 17 , .432 ) %- unsat
( 18 , .614 ) %- unsat
( 19 , .756 ) %- unsat
( 20 , .982 ) %- unsat
( 21 , 1.282 ) %- unsat
( 22 , 1.920 ) %- unsat
( 23 , 2.247 ) %- unsat
( 24 , 3.349 ) %- unsat
( 25 , 5.431 ) %- unsat
( 26 , 5.761 ) %- unsat
( 27 , 7.423 ) %- unsat
( 28 , 9.554 ) %- unsat
( 29 , 11.885 ) %- unsat
( 30 , 16.923 ) %- unsat
};\addlegendentry{$\pi_{+}$}
\end{axis}
\end{tikzpicture}

\iffalse
% property 3 , from graph 15 time is greater than 180 min 
% property 5 , all graphs unsat
%

\fi

%% file: conclusion.tex
We discussed the use of SMT-solvers for verifying dynamic network emulators described as many sorted first order logic formulas whose consistency can be checked using one of such solvers; in our work, we used z3. Once a network is implemented, its static instance can be verified at run-time. We discussed a possible verification solution and proposed an algorithm for checking that the instance does not violate the initial formula. Experimental results confirmed that the verification is the most efficient in an incremental \emph{link-by-link} way. For future work, we consider studying formula optimization techniques for decreasing the verification time.  We plan to consider more dynamic network parameters and the dependencies between them. We expect that such dependencies can be also verified using the proposed approach.